%% `Halley.tex': The phase of the scattering matrix
%%  by Jose M. Gracia-Bondia
%%  (16 March 2000)

\magnification\magstep1
\scrollmode

%%%%%%%%%%%%%%%%%%%%%
%% 1. Font management
%%%%%%%%%%%%%%%%%%%%%

%%% Not needed if there are no XyPic diagrams:
%%% \input xy       %% Load Xy-pic kernel
%%% \xyoption{all}  %% Extensions and features

\font\eightrm=cmr8                %% for the abstract
              %% for the authors

\font\tenbbb=msbm10  \font\sevenbbb=msbm7  \font\fivebbb=msbm5
\skewchar\tenbbb=127 \skewchar\sevenbbb=127 \skewchar\fivebbb=127
\newfam\bbbfam                    %% \fam 8
\textfont\bbbfam=\tenbbb \scriptfont\bbbfam=\sevenbbb
\scriptscriptfont\bbbfam=\fivebbb
      %% Blackboard bold (math mode only)

\font\tenams=msam10 \font\sevenams=msam7 %% more AMS symbols
\font\fiveams=msam5 \skewchar\tenams=127
\skewchar\sevenams=127 \skewchar\fiveams=127
\newfam\amsfam                    %% \fam 9
\textfont\amsfam=\tenams \scriptfont\amsfam=\sevenams
\scriptscriptfont\amsfam=\fiveams
                 %% small capitals

\mathchardef\square="2903         %% blank square box

%\MathPiTwofrak{eufm10}{10pt}{7pt}{5pt}  %% german fraktur letters

%%%%%%%%%%%%%%%%%%%%%%%
%% 2. Formatting macros
%%%%%%%%%%%%%%%%%%%%%%%

\def\cite#1{{\rm[#1]}}            %% reference citation
\def\eq#1{{\rm(#1)}}              %% print equation number
\def\opname#1{\mathop{\rm#1}\nolimits} %% operator name
\def\refno#1. #2\par{\smallskip\item{\rm[#1]}#2\par} %% print reference

%%%%%%%%%%%%%%%%%%%%%%%%%%%%%%%%%%%%%
%% 3. Sections and special paragraphs
%%%%%%%%%%%%%%%%%%%%%%%%%%%%%%%%%%%%%

\newcount\citenum
\newtoks\secnum
\def\computesecnum#1. #2\halt{\global\secnum={#1}}

\outer\def\section#1{\bigskip        %% section heading
      \computesecnum#1\halt          %% (stores section number)
      \leftline{\bf#1}\nobreak
      \smallskip\noindent\ignorespaces}

\outer\def\sectionstar#1{\bigskip    %% unnumbered section
      \leftline{\bf#1}\nobreak
      \smallskip\noindent\ignorespaces}

 %% \demo{Proof}
\def\enddemo{\ifmmode \ifinner \badenddemo \else \eqno\qed \fi
              \else \qed\par\smallskip \fi}  %% end-proof marker
              %% linebreak in mid-paragraph
         %% items with separation
  %% itemitems with separation
        %% begin new page

%%%%%%%%%%%%%%%%%%%%%%%%%%%%%
%% 4. Some internal machinery
%%%%%%%%%%%%%%%%%%%%%%%%%%%%%

\newbox\ncintdbox \newbox\ncinttbox %% noncommutative integral symbols
\setbox0=\hbox{$-$}
\setbox2=\hbox{$\displaystyle\int$}
\setbox\ncintdbox=\hbox{\rlap{\hbox
    to \wd2{\hskip-.125em\box2\relax\hfil}}\box0\kern.1em}
\setbox0=\hbox{$\vcenter{\hrule width 4pt}$}
\setbox2=\hbox{$\textstyle\int$}
\setbox\ncinttbox=\hbox{\rlap{\hbox
    to \wd2{\hskip-.125em\box2\relax\hfil}}\box0\kern.1em}

\def\qed{\allowbreak\qquad\null\nobreak\hfill\hbox{$\square$}}
\def\stroke{\mathbin|}            %% (for `\bbraket' and such)
\def\tribar{|\mkern-2mu|\mkern-2mu|} %% norm bars: |||

%%%%%%%%%%%%%%%%%%%%%%%%%%
%% 5. Simple abbreviations
%%%%%%%%%%%%%%%%%%%%%%%%%%

%%% Greek letters:
\def\a{\alpha}                    %% abbreviation for  \alpha
\def\b{\beta}                     %% abbreviation for  \beta
\def\dl{\delta}                   %% abbreviation for  \delta
                   %% abbreviation for  \Delta
\def\eps{\varepsilon}             %% abbreviation for  \varepsilon
                   %% abbreviation for  \Gamma
\def\ga{\gamma}                   %% abbreviation for  \gamma
                   %% abbreviation for  \kappa
\def\La{\Lambda}                  %% abbreviation for  \Lambda
\def\la{\lambda}                  %% abbreviation for  \lambda
                   %% abbreviation for  \Omega
                   %% abbreviation for  \omega
\def\sg{\sigma}                   %% abbreviation for  \sigma
                   %% abbreviation for  \Theta
\def\th{\theta}                   %% abbreviation for  \theta
\def\vf{\varphi}                  %% abbreviation for  \varphi

%%% Script letters:
                  %% an algebra
                  %% another algebra
                  %% de Rham currents
                  %% a projective module
                  %% Fourier transformation
                  %% Gelfand transformation
\def\H{{\cal H}}                  %% Hilbert space
                  %% a space of complex structures
                  %% compact operators, GLS symbols
                  %% Lie derivative
                  %% Moyal multiplier algebra
                  %% semigroup of projective modules
                 %% representative functions
                 %% symmetric algebra, Schwartz space
                %% Toeplitz algebra

%%% Blackboard boldface letters:
                 %% skewsymmetrization operator
                  %% complex numbers
                %% half-Dirac operator (`\eth')
                 %% quaternions
                  %% nonnegative integers
                  %% rational numbers
                  %% real numbers
                 %% sphere
                  %% circle as a group
                  %% integers

%%% Uppercase boldface math letters:
\def\Hb{{\bf H}}                  %% quantized free Hamiltonian
                  %% quantized number operator
                  %% quantized charge
\def\Sb{{\bf S}}                  %% quantum S-matrix
\def\Ub{{\bf U}}                  %% quantized propagator
\def\Vb{{\bf V}}                  %% quantized Hamiltonian

%%% Operator names:
\def\det{\opname{det}}            %% determinant (nonPlain)
\def\Ppart{\opname{P}}            %% principal-part distribution
\def\sign{\opname{sign}}          %% sign function
\def\Tm{\opname{T}}               %% time-ordered product
\def\Tr{\opname{Tr}}              %% trace of operator
\def\tr{\opname{tr}}              %% trace of matrix

%%% Roman letters:
               %% advanced propagator
\def\even{{\rm even}}             %% even degree forms
\def\odd{{\rm odd}}               %% odd degree forms
               %% retarded propagator

%%% Text abbreviations:
\def\wrt{with respect to}         %% abbreviation; usage: \wrt~

%%% Invisible super/subscripts:
\def\0{{\vphantom{\dagger}}}      %% invisible dagger (for alignment)

%%% Alternative symbol names:
\def\del{\partial}                %% abbreviation for  \partial
\def\op{\oplus}                   %% direct sum
                  %% tensor product
                    %% exterior product
\def\x{\times}                    %% cartesian product or cross
\def\7{\dagger}                   %% abbreviation for + symbol
\def\8{\bullet}                   %% anonymous degree
\def\.{\cdot}                     %% anonymous variable
\def\:{\colon}                    %% colon in  f: A -> B

%%%%%%%%%%%%%%%%%%%%%%
%% 6. Compound symbols
%%%%%%%%%%%%%%%%%%%%%%

%%% Symbol combinations:
                  %% derived spin representation
           %% left-handed spinor
        %% pair of chiral spinors
           %% right-handed spinor
         %% T-product, external legs
         %% T-product, vacuum loops

%%% Slashed symbols:
\def\Aslash{A\mkern-10mu/\,}      %% \ga^\mu A_\mu
 %% \ga^\mu \del_\mu
    %% generalized Dirac operator
       %% \ga^\mu k_\mu
\def\pslash{{p\mkern-8mu/}}       %% \ga^\mu p_\mu
\def\qslash{{q\mkern-8mu/}}       %% \ga^\mu q_\mu

%%% Repeated relations:
    %% repeated direct sum
  %% repeated tensor product
      %% repeated exterior product

%%% Arrows with riders:
 %% right arrow in diagram

%%% Large operators:
            %% double integral
 %% NC integral
\def\PVint{\,{\rm P}\!\!\int} %% principal value integral
 %% small sum-opr in display

%%% Small fractions in displays:
  %% tiny fraction  1/2
\def\thalf{{\textstyle{1\over2}}}    %% small fraction  1/2
   %% small fraction  i/2
 %% small fraction  i/4
 %% small fraction  1/4

%% Vacuum sector:
\def\vacin{0_{\rm in}}            %% in-vacuum vector
\def\vacout{0_{\rm out}}          %% out-vacuum vector
\def\vacpersamp{\<\vacin,\vacout>} %% persistence ampl

%%%%%%%%%%%%%%%%%%%%%%%%%%%%%
%% 7. Commands with arguments
%%%%%%%%%%%%%%%%%%%%%%%%%%%%%

%%% Accent-like macros:
\def\Bar#1{\overline{#1}}         %% closure operator
\def\Hat#1{\widehat{#1}}          %% abbreviation for  \widehat
\def\Onda#1{\widetilde{#1}}       %% abbreviation for \widetilde

%%% Emphasize words or phrases in text:
\def\textbf#1{{\bf#1}}            %% boldface phrase in text
\def\textit#1{{\it#1\/}}          %% italic phrase in text
\def\textrm#1{{\rm#1}}            %% roman phrase in italic text

%%% Separate words in displays:
\def\sepword#1{\qquad\hbox{#1}\quad} %% well-spaced words

%%% Enclose one argument with delimiters:
          %% bra vector
          %% ket vector
            %% set notation
\def\snorm#1{\mathopen{\tribar}{#1}\mathclose{\tribar}} %% norm |||x|||
\def\wick#1{\mathopen:#1\mathclose:} %% Wick-ordered operator

%%% Bilinear enclosures:
 %% rank one operator
\def\<#1,#2>{\langle#1\stroke#2\rangle} %% inner product (Dirac not'n)

%%% Fractions and derivatives:
     %% binomial coefficient
\def\dd#1{{\partial\over\partial#1}} %% partial derivation
\def\ddeval#1#2{\dd{#1}\biggr|_{#1=#2}} %% partial deriv at s = t
 %% derivative at 0
\def\frac#1#2{{#1\over#2}}        %% large fraction
 %% partial derivative
 %% any small fraction  a/b

%%% Matrices:
 %% 2 x 2 matrix
\def\twobytwoeven#1#2{\pmatrix{#1&0\cr0&#2\cr}} %% 2 x 2 diag matrix
\def\twobytwoodd#1#2{\pmatrix{0&#1\cr#2&0\cr}} %% 2 x 2 offdiag matrix

%%% Lists with dots:
   %% list:  a_1,...,a_n

%%%%%%%%%%%%%%%%%%%%%%%%%%%%
%% 8. Hyphenation exceptions
%%%%%%%%%%%%%%%%%%%%%%%%%%%%

\hyphenation{ap-pen-dix as-ymp-tot-ic Bo-go-liu-bov cor-res-pond-ence
cor-res-pond-ing de-riv-a-tive Eng-lish equi-va-lence equi-va-lent
equi-vari-ant Euler-ian Gauss-ian ge-ne-ral ge-ne-rate ge-ne-ra-ted
ge-o-des-ic geo-met-ric geo-met-ries geo-met-ry Hamil-ton-ian
Her-mit-ian ho-lo-no-my ideals in-fin-ite-ly in-fin-i-tes-i-mal
Lip-schitz neigh-bour-hood ope-ra-tor ope-ra-tors or-tho-go-nal
pa-ram-e-trize pro-duct pseu-do-dif-fer-en-tial qua-drat-ic
rec-tan-gle Rie-mann-ian semi-def-i-nite Sep-tem-ber skew-sym-met-ric
sum-ma-bi-li-ty sum-ma-ble to-po-lo-gi-cal to-po-lo-gy va-cuum}

%%%%%%%%%%%%%%%%%%%%
%% 9. Reference keys
%%%%%%%%%%%%%%%%%%%%

\def\reflist#1{\def\key##1{\advance\citenum by 1
                  \edef##1{\the\citenum}}#1} %% assign citation nums

\reflist{
\key\ConnesKr \key\Filk \key\Atlas \key\MinwallaRS \key\BrunettiF
\key\Grigore \key\HurthS \key\Pinter \key\EpsteinG \key\Philippe
\key\Scharf \key\Polaris \key\Rhea \key\ShaleS \key\BogoliubovS
\key\GrosseL \key\Ottesen \key\LangmannCocycle \key\Aldebaran
\key\Scheck \key\Estrada \key\BerestetskiiLP \key\ItzyksonZ
\key\GreinerQED \key\LangmannM \key\JauchR \key\Ruijsenaars
\key\ScharfW \key\Bellissard}

%%% Document begins here:

\centerline{\bf THE PHASE OF THE  SCATTERING MATRIX}
\bigskip
\centerline{Jos\'e M. Gracia-Bond\'{\i}a$^{1,2}$}
\medskip
\centerline{\it $^1$Department of Theoretical Physics I, University
Complutense, Madrid 28040, SPAIN}
\smallskip
\centerline{\it $^2$Department of Theoretical Physics, University of
Zaragoza, Zaragoza 50009, SPAIN}

\bigskip
%% Abstract
{\narrower\eightrm\baselineskip=9.5pt
\noindent
Vacuum polarization in external fields is treated by way of calculating
---exactly and then perturbatively--- the phase of the quantum
scattering matrix in the Shale--Stinespring approach to field
theory. The link between the Shale--Stinespring  method and the
Epstein--Glaser renormalization procedure is highlighted.
\par}

\bigskip

\section{1. Introduction}

Renormalization theory is bound to suffer a reappraisal in the light
of the reconstruction of Zimmermann's forest formula in Hopf-algebraic
terms, together with the interpretation of the dimensional
regularization method given by Connes and Kreimer~\cite{\ConnesKr},
and the birth of quantum field theory on noncommutative
spaces~\cite{\Filk,~\Atlas}, together with the evidence that
Yang--Mills theories on noncommutative manifolds are ultraviolet
divergent ---see~\cite{\MinwallaRS} and the other references in that
paper.

Now, part of the advantages and new popularity (see, for instance,
\cite{\BrunettiF--\Pinter})
%%%\cite{\BrunettiF,\Grigore,\HurthS,\Pinter}
of the Epstein--Glaser renormalization method
\cite{~\EpsteinG,\Philippe} stems from the fact that it is locally
defined, and so in principle applies to models on nonflat manifolds.
There is however some contention on whether, as claimed by some
practitioners~\cite{\Scharf}, the Epstein--Glaser method is a
fundamental one.

In cases like these, it sometimes helps to look at a simpler problem,
for which an absolutely reliable method is known, to see how the
marketed procedures fare in its respect.

The chosen problem is that of vacuum polarization in external fields
and the chosen reliable method is the Shale--Stinespring approach to
linear quantum field theories. This is an entirely rigorous algebraic
method; in~\cite{\Atlas} its \textit{prima facie} applicability to
``implementable'' theories on noncommutative as well as commutative
spaces was brought to the fore.

After reviewing the Shale-Stinespring theorem in Section~2, in the
body of this article we show, by a refinement of its technique, that
the phase of the scattering matrix is well defined and finite for
implementable linear theories. We give explicit formulae for the
phase.

We then exploit in QED a perturbative version of this approach, which
leads in a direct way to the formulas reached by Scharf~\cite{\Scharf}
in his account of vacuum polarization in QED by the Epstein--Glaser
method. Some simplification of his calculations results from using
gauge-invariant variables. The contention by Scharf and followers that
the Epstein--Glaser renormalization procedure is a fundamental one is
vindicated to some extent.

\section{2. A reminder on Shale--Stinespring theory}

Let $\H = \H^+ \op \H^- =: P_+\H \op P_-\H$ be a Hilbert space, graded
by the projections $P_{\pm}$ on the positive and negative spectral
subspaces for a free Dirac operator. Operators on $\H$ are presented
in block form:
$$
A = \pmatrix{A_{++} & A_{+-} \cr A_{-+} & A_{--} \cr}.
$$
We have in mind particularly the classical (or ``first-quantized'')
scattering matrix
$$
S = \pmatrix{S_{++} & S_{+-} \cr S_{-+} & S_{--} \cr}.
$$
Introduce a nomenclature for its even and odd parts:
$$
p_S := S_\even = \twobytwoeven{S_{++}}{S_{--}}, \quad
q_S := S_\odd  = \twobytwoodd{S_{+-}}{S_{-+}}.
$$
Similarly,
$$
S^{-1} = S^\7
 = \pmatrix{S_{++}^\7 & S_{-+}^\7 \cr S_{+-}^\7 & S_{--}^\7 \cr},
\quad p_{S^\7} = \twobytwoeven{S_{++}^\7}{S_{--}^\7},
\quad q_{S^\7} = \twobytwoodd{S_{-+}^\7}{S_{+-}^\7}.
$$
Unitarity of $S$ gives the identities
$$
\eqalign{
S_{++}S_{++}^\7 + S_{+-}S_{+-}^\7
&= S_{++}^\7S_{++} + S_{-+}^\7S_{-+} = P_+,
\cr
S_{--}S_{--}^\7 + S_{-+}S_{-+}^\7
&= S_{--}^\7S_{--} + S_{+-}^\7S_{+-} = P_-,
\cr
S_{++}S_{-+}^\7 + S_{+-}S_{--}^\7
&= S_{++}^\7S_{+-} + S_{-+}^\7S_{--} = 0,
\cr
S_{--}S_{+-}^\7 + S_{-+}S_{++}^\7
&= S_{--}^\7S_{-+} + S_{+-}^\7S_{++} = 0.
\cr}
$$
It is clear that $p_S^{-1}$ exists if and only if $S_{++}$ and
$S_{--}$ are invertible as operators on $\H^+$ and on~$\H^-$,
respectively; this is the generic case, that always holds when $S$ is
close to the identity, and will be the only one considered in the
sequel. We then define the skewadjoint operators
$$
T_S := q_S p_S^{-1}
 = \twobytwoodd{S_{+-}S_{--}^{-1}}{S_{-+}S_{++}^{-1}}, \quad
\Hat T_S := T_{S^\7}
 = \twobytwoodd{-S_{++}^{-1}S_{+-}}{-S_{--}^{-1}S_{-+}}.
$$

Consider the Fock space constructed on $\H$ with the new scalar product
$$
\<\eta, \vf> := \<\eta_+, \vf_+> + \<\vf_-, \eta_->,
\eqno (1)
$$
where $\eta_{\pm} := P_{\pm}\eta$. Let $\{\phi_k\}$ and $\{\psi_k\}$
denote arbitrary orthonormal bases for $\H^+$ and~$\H^-$,
respectively; we shall abbreviate $b_k := b(\phi_k)$,
$d_k := d(\psi_k)$ in the notation of the ``particle'' and
``antiparticle'' annihilation operators, and similarly for the
creation operators $b_k^\7$, $d_k^\7$. For any operator $A$ on $\H$ we
have the quantum (or ``second-quantized'') counterpart, acting on Fock
space:
$$
d\La(A) := b^\7 A_{++} b + b^\7 A_{+-} d^\7 + d A_{-+} b
  + \wick{d A_{--} d^\7}\,.
$$
Here, for instance, $\wick{d A_{--} d^\7}$ denotes
$-\sum_{j,k} d_k^\7 \<\psi_j, A_{--}\psi_k> d_j$, the double colon
meaning, as usual, a normally ordered product, and $b^\7 A_{+-} d^\7$
is $\sum_{j,k} b_k^\7 \<\phi_k, A_{+-}\psi_j> d_j^\7$; the other cases
should be clear.

This rule corresponds to the infinitesimal spin
representation~\cite{\Polaris}; it is independent of the drafted
orthonormal bases and makes sense only when $A_{+-}$, $A_{+-}$ are
Hilbert--Schmidt.

The rule is mainly applied to selfadjoint operators, and yields (at
least formally) selfadjoint operators in turn. For instance, in QED
the free Dirac equation is written as
$$
i\dd{t}\psi = \b m\psi - i \vec\a \dd{\vec x}\psi =: D_0\psi,
$$
where the Dirac matrices, say in the chiral representation, are given
by
$$
\vec\a := \ga^0\vec\ga = \twobytwoeven{\vec\sg}{-\vec\sg},  \quad
 \b := \ga^0 := \twobytwoodd{1_{2}}{1_{2}}.
$$
To this \textit{classical} selfadjoint operator $D_0$ corresponds the
\textit{quantum} free Hamiltonian
$$
\Hb_0 := d\La(D_0) = b^\7 D_0 b^\0 + \wick{d D_0 d^\7}\,,
$$
which is a positive operator. The \textit{classical} interaction
Hamiltonian is of the form
$$
H(t) = e(A^0(t) - \vec\a\.\vec A(t)),
$$
where $e$ denotes the electromagnetic coupling constant and
$(A^0,\vec A) =: A$ the electromagnetic vector potential, a real
$c$-number function. Note the covariant form $H(t) = e \ga^0 \Aslash$,
with $\Aslash := \ga^\mu A_\mu$. In the interaction picture one
considers
$$
V(t) := e^{iD_0t} H(t) e^{-iD_0t}.
\eqno (2)
$$
The \textit{quantum} interaction Hamiltonian is then
$$
\Vb(t) := d\La(V(t)) = b^\7 V_{++}(t) b^\0 + b^\7 V_{+-}(t) d^\7
 + d^\0 V_{-+}(t) b^\0 + \wick{d  V_{--}(t) d^\7}\,.
\eqno (3a)
$$
This can be rewritten in terms of the formal fermion field $\Psi$, as
$$
\Vb(t) = \int d^3x \,\wick{\Bar\Psi(x) \Aslash(x) \Psi(x)}\,,
\eqno (3b)
$$
with $x = (t,\vec x)$ and the bar meaning the Dirac adjoint. For that,
just write the fermion field in the form
$$
\Psi(x) = \sum_k(b_k \phi_k(x) + d_k^\7 \psi_k(x)),
$$
where $\phi_k(x) = e^{-iD_0t} \phi_k(\vec x)$, and similarly for the
$\psi$'s.

\smallskip

For the quantum scattering matrix $\Sb$ we need instead the global
spin representation~\cite{\Polaris,\Rhea}, which we call $\La$. It is
given by
$$
\Sb := e^{i\th}\La(S) = e^{i\th}|\vacpersamp|\, \wick{\exp d\La(I)}
 = \vacpersamp\, \wick{\exp d\La(I)}\,.
$$
Here $\vacin$ denotes the incoming vacuum, $\vacout := \Sb\,\vacin$,
and
$$
I := \pmatrix{(S_{++}^\7)^{-1} - 1 & S_{+-}^\0 S_{--}^{-1} \cr
              S_{--}^{-1} S_{-+}^\0 & 1 - S_{--}^{-1} \cr}.
$$
Again, this makes sense if and only if $S_{+-}$, $S_{-+}$ are
Hilbert--Schmidt, and then we say that $S$ is \textit{implementable}.
The absolute value of the vacuum persistence amplitude $\vacpersamp$
is given by
$$
\eqalign{
|\vacpersamp|
&= \det^{-1/4}(1 - T_S^2) = \det^{1/2}(S_{--}^\0 S_{--}^\7)
 = \det^{1/2}(S_{++}^\0 S_{++}^\7)
\cr
&= \det^{1/2}(1 - S_{+-}^\0 S_{+-}^\7)
 = \det^{1/2}(1 - S_{-+}^\0 S_{-+}^\7).
\cr}
$$
For our present purposes, this is the content of the
Shale--Stinespring theorem~\cite{\Polaris,~\ShaleS}. The phase $\th$
is in principle undetermined and conventionally taken equal to zero;
this is all that is needed to compute transition probabilities.

The phase of the vacuum persistence amplitude \textit{does} matter
physically, however: the current density is modified with respect to
the free field situation by the vacuum polarization effect (that bears
on the radiative correction to the photon propagator in the nonlinear
theory), and the interacting current density is found by functional
derivation of $\Sb$ with respect to the gauge potential, in which the
phase intervenes~\cite{\BogoliubovS}.

The question is then to find an appropriate and computable definition
for the phase of the quantum scattering matrix.

\section{3. Computing the phase in the Shale--Stinespring framework}

The difficulty comes from the fact that the global spin representation
is projective. Let $U(s,t)$ be the classical unitary propagator in the
interaction representation, which interpolates between the identity
and $S$. Then $U(s,t)$ solves the equation
$$
U(s,t) = 1 - i\int_t^s V(u) U(u,t) \,du;
\eqno (4)
$$
an explicit form being given by the Dyson expansion
$$
U(s,t) = 1 + \sum_{n=1}^\infty (-i)^n \int_t^s V(t_1) \int_t^{t_1}
 V(t_2) \cdots \int_t^{t_{n-1}} V(t_n) \,dt_n \dots dt_2 \,dt_1,
$$
from which follow the propagator properties:
$$
U(t,t) = 1, \qquad U(t,s) U(s,r) = U(t,r).
$$
Here $S = U(\infty,-\infty)$.

Now, for unitary implementable operators $U_1$, $U_2$, in general
$$
\La(U_1) \La(U_2) = c(U_1,U_2) \La(U_1U_2),
$$
where the cocycle $c$ (with the vanishing phase convention) is
given~\cite{\Polaris,~\Rhea} by
$$
c(U_1,U_2) = \exp\bigl(i\arg\det^{1/2}(1 - T_{U_2}\Hat T_{U_1})\bigr)
= \exp\bigl(i\arg\det^{1/2}(p^{-1}_{U_1}p_{U_1U_2}p^{-1}_{U_2})\bigr).
\eqno (5)
$$
This is not a trivial cocycle because the determinants of the $p_U$
operators are not individually defined in general. Assume that the
interpolating family $U(s,t)$ is implementable and (strongly)
differentiable \wrt\ its parameters ---this will happen if the
external field is sufficiently well behaved. Then
$$
\La(U(s,t)) \La(U(t,r)) = c(s,t,r) \La(U(s,r)),
\eqno (6)
$$
with an obvious notation. Still, $c(t,t,r) = c(s,t,t) = 1$. On
the other hand, it can be shown~\cite{\Polaris} that
$$
i\ddeval{s}{t} \La(U(s,t)) = \Vb(t).
$$

We now seek to redefine $\La(U(s,t))$ by multiplying a phase factor
$e^{i\th(s,t)}$ so that the new quantum family
$\Ub(s,t) := e^{i\th(s,t)}\La(U(s,t))$ fulfils
$$
\Ub(t,t) = 1; \quad \Ub(s,t)\Ub(t,r) = \Ub(s,r),
\eqno (7)
$$
just like the classical propagator. If we manage that, then
$\th(+\infty,-\infty)$ will have every right to be called the
\textit{phase} of the quantum scattering operator $\Sb$. Let
$c(s,t,r) =: \exp(i\xi(s,t,r))$. Differentiating equation~\eq{6},
one gets:
$$
\Vb(t) \La(U(t,r))
 = i\ddeval{s}{t} c(s,t,r) \La(U(t,r)) + i\dd{t}\La(U(t,r)).
$$
Then, we redefine
$$
\Ub(s,t) := \exp\biggl( i\int_t^s \ddeval{\la}{\tau} \xi(\la,\tau,t)
 \,d\tau\biggr) \La(U(s,t)),
$$
which clearly satisfies
$$
\Ub(s,t) = 1 - i\int_t^s \Vb(u)\Ub(u,s) \,du.
$$
This equation is the quantized version of~\eq{4} and sports the same
kind of iteration solution:
$$
\Ub(s,t) = 1 + \sum_{n=1}^\infty (-i)^n \int_t^s \Vb(t_1) \int_t^{t_1}
 \Vb(t_2) \cdots \int_t^{t_{n-1}} \Vb(t_n) \,dt_n \dots dt_2 \,dt_1.
\eqno (8)
$$
In the present case, the quantum Dyson expansion \textit{is} rigorous:
although $\Vb$ is an unbounded operator, it is a pretty tame one. Let
$E_m$ denote the projector on states containing at most $m$ particles.
Then, in view of~\eq{3}, $\Vb E_m$ is a bounded operator from $E_m$
into $E_{m+2}$, and the norms of
$$
\Ub(s,t)_{:n}\,E_m
 := \int_s^t \Vb(t_1) \int_s^{t_1} \Vb(t_2) \cdots \int_s^{t_{n-1}}
      \Vb(t_n) \,dt_n \dots dt_2 \,dt_1\, E_m
$$
can be easily estimated. To see that, one introduces the norm
$$
\snorm{V(t)} := \|V_\even(t)\| + \|V_\odd(t)\|_2,
$$
where the latter is the Hilbert--Schmidt norm. By continuity and
uniform boundedness, $\snorm{V(t)} \leq a(s,r)$ holds for some finite
function $a(s,r)$, when $r \leq t \leq s$, and is not difficult to
check that there are constants $C_m$ such that
$$
\|\Vb(t) E_m\| \leq C_m\,a(s,r).
$$
Consult \cite{\GrosseL,~\Ottesen} for precise analyses of these bounds
(encompassing also the boson case). On the other hand, from the
integral equation~\eq{4},
$$
\snorm{U(s,r)} \leq 1 + (s - r)\,a(s,r).
$$
Putting both inequalities together, one gets the estimate
$$
\|\Ub(s,t)_{:n}\,E_m\|
 \leq C_m C_{m+2}\dots C_{m+2n-2} \frac{(s-t)^n a(s,t)^n}{n!},
$$
with the result that the series in~\eq{8} indeed converges to a
unitary operator, for $s - t$ small enough. Equation~\eq{7} then
follows from~\eq{8} and allow us to extend the validity of the last
conclusion---and, in turn, its own domain of validity. More detail on
this is found in the important paper~\cite{\LangmannCocycle}.

We can give an exact formula for the phase now. Use the standard
identities
$$
\frac{d}{dt}(\arg z(t)) = \Im \frac{d}{dt}(\log z(t)),  \qquad
\frac{d}{dt}(\log\det A(t))
 = \Tr\biggl( A(t)^{-1}\,\frac{dA(t)}{dt} \biggr),
$$
which give, from~\eq{5},
$$
\ddeval{\la}{\tau} \,\xi(\la,\tau,t)
 = -\frac{1}{2} \Im \Tr\biggl( T_{U(\tau,t)}
   \ddeval{\la}{\tau} \Hat T_{U(\la,\tau)} \biggr).
$$
Therefore,
$$
\th(s,t) = -\frac{1}{2} \Im \int_t^s \Tr\biggl( T_{U(\tau,t)}
    \ddeval{\la}{\tau} \Hat T_{U(\la,\tau)} \biggr)\,d\tau
 = -\frac{i}{4} \int_t^s \Tr\biggl[ \ddeval{\la}{\tau}
    \Hat T_{U(\la,\tau)}, T_{U(\tau,t)}\biggr] \,d\tau.
$$
(The last expression is more symmetrical; the trace of this commutator
is \textit{not} zero, because it is taken in Fock space, whereupon, in
view of the form of the scalar product~\eq{1},
$$
\Tr\pmatrix{A_{++} & A_{+-} \cr A_{-+} & A_{--}}
 = \Tr A_{++} + \Tr A_{--}^\7 \,.)
$$
The phase of the scattering matrix is then
$$
\th = -\frac{1}{2} \Im \int_{-\infty}^\infty
 \Tr\biggl( T_{U(\tau,-\infty)} \ddeval{\la}{\tau}
  \Hat T_{U(\la,\tau)} \biggr) \,d\tau.
\eqno (9)
$$
The analogous formula, with the same notation, for the boson case was
first given, to the best of our knowledge, by V\'arilly and the
author~\cite{\Aldebaran}; it differs only by a sign. Then, equivalent
formulae both for the boson and fermion cases were found by
Langmann~\cite{\LangmannCocycle}. The latter apply to charged fields,
which are the ones considered in this paper. However, under the
form~\eq{9} and with a suitable interpretation, the phase formula
is applicable to Majorana fields, which are more general than charged
fields~\cite{\Scheck}.

Also note, before continuing, that
$$
\ddeval{s}{t} \th(s,t) = \ddeval{\la}{s} \xi(\la,s,s) = 0.
\eqno (10)
$$
In other words, there is \textit{no} contribution from the coincidence
points of $\Hat T_{U(\la,\tau)}$ and $T_{U(\tau,t)}$. This will prove
to be the crucial remark.

One has simply
$\del/\del\la\bigr|_{\la=\tau} \hat T_{U(\la,\tau)} = iV_\odd(\tau)$
in our present framework. Therefore, on calling
$T(\tau,t) := T_{U(\tau,t)}$, finally:
$$
\th(s,t) = \frac{1}{2} \int_t^s \Tr(V_{+-}(\tau)T_{-+}(\tau,t)
 - T_{+-}(\tau,t) V_{-+}(\tau)) \,d\tau,
$$
a rather elegant expression. Note that it differs from zero only at
second order in perturbation theory. At that order,
$$
\th_{:2}(s,t) = \frac{1}{2} \int_t^s \Tr(V_{+-}(\tau)U_{-+:1}(\tau,t)
 - U_{+-:1}(\tau,t) V_{-+}(\tau)) \,d\tau
$$
with an obvious notation. Since
$U_{:1}(\tau,t) = -i\int_\tau^t V(\tau)\,d\tau$, we set out to compute
$$
\th_{:2}(s,t) = -\frac{i}{2} \int_s^t \Tr\bigl[ V_{+-}(\tau)
 \biggl( \int_\tau^t V(\la)\,d\la \biggr) -
 \biggl( \int_\tau^t V(\la)\,d\la \biggr) V_{-+}(\tau) \bigr] \,d\tau.
$$
Of course, \eq{10} still applies at this approximation. The total
phase $\th_{:2} := \th_{:2}(\infty,-\infty)$ at this approximation is
then
$$
\th_{:2} = -\frac{i}{2} \int_{-\infty}^\infty \int_{-\infty}^\infty
 \th(t_1 - t_2) (V_{+-}(t_1) V_{-+}(t_2) - V_{+-}(t_2) V_{-+}(t_1))
  \,dt_1 \,dt_2.
\eqno (11a)
$$
It should be clear that, at the same order of approximation, this is
precisely
$$
-\frac{1}{2} \Im \biggl< \vacin \biggm|
  \int_{-\infty}^\infty \int_{-\infty}^\infty
    \Tm[\Vb(t_1)\Vb(t_2)] \,dt_1 \,dt_2\ \vacin \biggr>,
\eqno (11b)
$$
where $\Tm$ denotes the time-ordered product.

\section{4. Perturbative calculation of the phase in quantum
electrodynamics}

In QED, from~\eq{2}, or working like in the derivation of~\eq{3b},
the integrals~\eq{11} are recast as
$$
\eqalignno{
\th_{:2} = \Im \frac{e^2}{2} \int &\th(t_1 - t_2)
 \tr[\Aslash(x_1) S^-(x_1 - x_2) \Aslash(x_2) S^+(x_2 - x_1)
\cr
&\qquad - \Aslash(x_1) S^+(x_1 - x_2) \Aslash(x_2) S^-(x_2 - x_1)]
  \,d^4x_1 \,d^4x_2.
& (12a) \cr}
$$
Here $S^{\pm}$ denote the Wightman ``functions''.

The first thing to remark is that, since
$$
S^-(x) \ga^\nu S^+(-x) - S^+(x) \ga^\nu S^-(-x)
 = S_{JP}(x) \ga^\nu S^+(-x) - S^+(x) \ga^\nu S_{JP}(-x),
$$
and $S_{JP}$ has support inside the lightcone, then the integrand has
support inside the lightcone. That allows one to substitute for
$\th(t_1 - t_2)$ the covariant expression
$\th((v(x_1 - x_2))) =: \chi(x_1 - x_2)$, where $v$ is an
\textit{arbitrary} timelike vector, which can thus be varied at will,
and a parenthesis has been used to denote the Minkowski product
$g_{\mu\nu}y^\mu x^\nu=:(yx)$ of two four-vectors $y,x$. Let then
$$
\eqalign{
\Onda F^{\mu\nu}(x) &:= \tr[\ga^\mu S^-(x) \ga^\nu S^+(-x)
    - \ga^\mu S^+(x) \ga^\nu S^-(-x)],
\cr
F^{\mu\nu}(x) &:= \chi(x) \Onda F^{\mu\nu}(x).
\cr}
$$

The second thing to remark is that the last expression ---and
hence~\eq{12a}--- is only formal: $\chi\Onda F^{\mu\nu}$ is actually
undefined as a product of distributions in view of the singularities
of $\Onda F^{\mu\nu}$ on the lightcone. Because the apparent trouble
occurs at the coincidence points and since
$\chi(x)\Onda F^{\mu\nu}(x)$ makes sense for $x \neq 0$, one can try
to define $\chi\Onda F^{\mu\nu}$ as a distributional
\textit{extension} ---or ``regularization'' in the terminology
of~\cite{\Estrada}--- of the latter. The scaling
degree~\cite{\BrunettiF} or singular order~\cite{\Scharf} of the
integral of the product of the two Wightman functions is~2; therefore,
distinct extensions of this quantity will differ by linear
combinations of the delta function at the origin and its derivatives
up to order two ---i.e., by polynomials in~$k$ of degree at most two
in momentum space. The procedure is undoubtedly sound in the present
case, as we know \textit{a priori} the phase to be finite. Moreover,
our framework will allow to select the ``good'' extension.

It is indeed convenient to work in momentum space. There,
$$
\th_{:2} = \frac{e^2}{2} (2\pi)^2\, \Im \int F^{\nu\mu}(k) A_\mu(k)
 \Bar A_\nu(k) \,d^4k,
\eqno (12b)
$$
where we take into account that $A(-k) = \Bar A(k)$, with the bar
meaning here complex conjugation, because $A(x)$ is real; and,
formally, $F^{\mu\nu}(k) = (2\pi)^{-1/2} \chi * \Onda F^{\mu\nu}(k)$
with $*$ denoting ordinary convolution. We recall that, for timelike
$k$ and when choosing a frame in which $k = (k^0,\vec0)$,
$$
\chi(k) = \frac{-i\dl(\vec k)}{\sqrt{2\pi}(k^0 - i\eps)}
 = \frac{-i\dl(\vec k)}{\sqrt{2\pi}} \biggl(
     \Ppart \frac{1}{k^0} + i\pi\dl(k^0) \biggr).
\eqno (13)
$$

To compute $\Onda F^{\mu\nu}(k)$, one looks at the Fourier transform
of $\tr[\ga^\mu S^+(x) \ga^\nu S^-(x)]$. By using the well known
expressions of the Wightman functions in momentum space, this is
expressed as
$$
\eqalignno{
-\frac1{(2\pi)^4} \int &\tr[\ga^\mu(\pslash + m) \ga^\nu(\qslash - m)]
    \th(p^0) \th(q^0) \dl(p^2 - m^2)
\cr
&\quad \x \dl(q^2 - m^2) \dl^4(k - p - q) \,d^4q \,d^4p
 =: -\frac1{(2\pi)^4} T^{\nu\mu}(k).
& (14) \cr}
$$
Moreover, $\tr[(\pslash + m)\ga^\mu(\qslash - m)\ga^\nu]
 = 4(p^\mu q^\nu + q^\mu p^\nu - ((pq) - m^2)g^{\mu\nu})$. Then one of
the integrations in~\eq{14} is immediately disposed of, with the help
of the $\dl^4$-function. The other is easily performed, with the help
of the remaining $\dl$-functions, again by choosing a frame in which
$k = (k^0,\vec 0)$ so that $\Onda F^{\mu\nu}$ can be regarded as a
function of only one variable; we obtain
$$
T^{\mu\nu}(k) = \frac{2\pi}{3}
 \biggl( \frac{k^\mu k^\nu}{k^2} - g^{\mu\nu} \biggr)
 \bigl[ k^2(1 + \ga(k^2)) (1 - 2\ga(k^2))^{1/2}
                  \th(1 - 2\ga(k^2)) \th(k^0) \bigr],
$$
where $\ga(k^2) := 2m^2/k^2$. All this is found in many books
\cite{\Scharf,~\BerestetskiiLP,~\ItzyksonZ}.

Thus we have been led formally to compute $\chi * \Onda F^{\mu\nu}$,
where $\Onda F^{\mu\nu}(k)$ equals
$$
\frac1{3(2\pi)^3} \biggl(\frac{k^\mu k^\nu}{k^2} - g^{\mu\nu}\biggr)
  \bigl[ k^2 (1 + \ga(k^2)) (1 - 2\ga(k^2))^{1/2} \th(1 - 2\ga(k^2))
         \sign(k^0) \bigr].
$$
That indeed behaves as a polynomial of degree two at high momentum
transfer, confirming that the singular degree of $F^{\mu\nu}$ is two.

The correct (unique) recipe to regularize the imaginary part of
$\chi * \Onda F^{\mu\nu}$ is selected by prescribing that the result
$F^{\mu\nu}$ vanishes, together with derivatives up to order two, at
zero momentum. This kills the delta function at the origin and its
derivatives in configuration space, which otherwise \textit{would
give a nonzero contribution} to $\del/\del s\bigr|_{s=t} \th(s,t)$,
contradicting~\eq{10}. It is clear now, from the $\dl$-function
in~\eq{13}, that
$$
\Re F^{\mu\nu}(k) = \thalf \Onda F^{\mu\nu}(k),
$$
whereas the relation between the real and imaginary parts of
$F^{\mu\nu}$ is then given by a subtracted (at the origin) dispersion
relation
$$
\Im F^{\mu\nu}(k^0)
= - \frac{(k^0)^3}{\pi} \PVint_{-\infty}^\infty
       \frac{\Re F^{\mu\nu}(\zeta)}{\zeta^3(\zeta - k^0)} \,d\zeta.
\eqno (15)
$$

The prescription that $F^{\mu\nu}$ possess a zero of the third
(indeed, fourth) order at $k = 0$ can be independently justified by an
argument that, although heuristic, we deem very strong in the present
context. On invoking the Maxwell equations (in the Lorentz gauge)
$A^\mu(k) = \jmath^\mu(k)/k^2$ to conjure up the source $\jmath$ of
the classical field, and on introducing
$$
G(k) := \frac{1}{k^2} (1 + \ga(k^2)) (1 - 2\ga(k^2))^{1/2}
  \th(1 - 2\ga(k^2)) \sign(k^0),
$$
one gets
$$
\frac{e^2}{2} (2\pi)^2 \Re \int F^{\mu\nu}(k) A_\mu(k) \Bar A_\nu(k)
  \,d^4k
 = - \frac{e^2}{24\pi} \int (\jmath(k)\bar\jmath(k)) G(k) \,d^4k.
\eqno (16)
$$
The continuity equation $(\jmath(k)k) = 0$ has been employed to
simplify the result. This simple expression exhibits only
gauge-invariant variables.

Now, we remark that (classical) gauge transformations are in general
\textit{not implementable} in $1+3$ dimensions. This is a very good
indicator of the existence of ultraviolet divergences in the nonlinear
theory, and indeed it was used by J.~C. V\'arilly and the author
in~\cite{\Atlas} to point out that QFT theories on noncommutative
manifolds had to be ultraviolet divergent. On the other hand, in the
linear theory selfinteraction is absent, so we would not expect
ultraviolet divergences on physical grounds. That nonimplementability
is the \textit{only} source of spurious divergence difficulties. We
therefore expect to be able to express the phase in terms of
gauge-invariant variables, in a similar way to \eq{16}:
$$
\th_{:2} = -\frac{e^2}{24\pi} \int (\jmath(k)\bar\jmath(k))H(k)\,d^4k,
\eqno (17a)
$$
with $H(k)$ is \textit{regular} at $k = 0$; this is equivalent to
$F^{\mu\nu}$ having the aforementioned behaviour at the origin in
momentum space.

Taking into account that $G$ is odd, equation~\eq{15} leads
immediately to a simple form for~$H$:
$$
H(k) = \frac{1}{\pi} \PVint_{4m^2}^\infty
  \frac{(1 + \ga(\la))(1 - 2\ga(\la))^{1/2}}{\la(\la - k^2)} \,d\la.
\eqno (17b)
$$
The restriction to timelike $k$ can be removed by analytic
continuation. Making the change of variable $\la =: 4m^2/(1 - v^2)$,
we get
$$
H(k) = \frac{3}{\pi k^2} \int_0^1
  \frac{v^2 - v^4/3}{v^2 - 1 + 4m^2/k^2}\,dv,
$$
which is essentially the expression one finds in
textbooks~\cite{\GreinerQED, pp.~249--252} for the vacuum polarization
functional, after renormalization. The last integral can be easily carried
out analytically, and it is then an instructive exercise to check that the
function $H$ is perfectly smooth at $k = 0$ (at $k^2 = 4m^2$, the
onset of the absorptive part, $H$ has a cusp). We shall not go into
the details.

\smallskip

Before rushing to the conclusions, a comment is in order: we have more
or less treated the $A(x)$ as test functions, guaranteeing
implementability of the interpolating operators, for the sake of the
argument. However, it is clear that the final formula for the
phase~\eq{9} is acceptable with only the milder requirement of the
implementability of the scattering operator; consult~\cite{\LangmannM}
for a very efficient removal of technical conditions on the
potentials, for this last purpose.

\section{5. Conclusions}

In this paper we have performed what amounts to an
\textit{ab initio} finite calculation of the ``bubble'' diagrams
in linear quantum field theory.

Now, at first significant order in QED this is essentially the same as
the one-loop vacuum polarization or ``photon self-energy'' diagram
(see in this respect~\cite{\JauchR, pp.~195--196}). The computation
done here does not appear to have been pushed to that finish line
before now, although the tools have been there since the seventies at
least~\cite{\Ruijsenaars}. (Besides
V\'arilly and the author~\cite{\Aldebaran,\Rhea}, Langmann and
Mickelsson~\cite{\LangmannCocycle,\LangmannM} came close in the
nineties.) The main point is that the ``local causality
condition''~\eq{10} selects the correct prescription among all the
(finite) regularizations, with recourse to neither heuristic
arguments~\cite{\Scharf} nor the extremely long and complicated
``nonperturbative proof'' in~\cite{\ScharfW}.

On the other hand, it will not have escaped the reader's attention
that, in order to avoid pitfalls, we reorganize the calculation in the
same way as~\cite{\Scharf}. The whole procedure is thus in the spirit
of the Epstein--Glaser renormalization procedure, where there are
Feynman graphs, but the Feynman rules do not necessarily apply ---we
avoided rewriting~\eq{12a} in terms of Feynman propagators. It is
remarkable that, in the boson case, the quantum scattering matrix was
found long ago by the Epstein--Glaser method by
Bellissard~\cite{\Bellissard} ---without the phase.

{}From formulae~\eq{16} and ~\eq{17} one can easily verify
\textit{a posteriori\/} Bogoliubov's causality condition
$$
\frac{\dl}{\dl A(x_1)} \biggl< \vacin \biggm|
 \Sb^\7 \frac{\dl\Sb}{\dl A(x_2)}\ \vacin \biggr> = 0
 \sepword{for} x_1^0 > x_2^0,
$$
which was the starting point of Epstein and Glaser.

\sectionstar{Acknowledgments}

I am happy to thank Ricardo Estrada, H\'ector Figueroa, Steve Fulling,
Carmelo P. Mart\'{\i}n and Joe V\'arilly for discussions, and my
students Esteban Araya and Pablo Blanco for help with drawing~$H$. The
warm hospitality of the departments of theoretical physics at the
Complutense in Madrid and at Zaragoza, and the financial
support of Sab\'aticos UCM 1999 are most gratefully acknowledged.

\sectionstar{References}

\frenchspacing

\refno\ConnesKr.
A. Connes and D. Kreimer,
``Renormalization in quantum field theory and the Rie\-mann--Hilbert
problem I: the Hopf algebra structure of graphs and the main theorem'',
hep-th/9912092, IHES, 1999.

\refno\Filk.
T. Filk,
``Divergences in a field theory on quantum space'',
Phys. Lett. B {\bf 376} (1996), 53--58.

\refno\Atlas.
J. C. V\'arilly and J. M. Gracia-Bond\'{\i}a,
``On the ultraviolet behaviour of quantum fields over noncommutative
manifolds'',
Int. J. Mod. Phys. A {\bf 14} (1999), 1305--1323.

\refno\MinwallaRS.
S. Minwalla, M. Van~Raamsdonk and N. Seiberg,
``Noncommutative perturbative dynamics'',
hep-th/9912072, Princeton, 1999.

\refno\BrunettiF.
R. Brunetti and K. Fredenhagen,
``Microlocal analysis and interacting quantum field theories:
renormalization on physical backgrounds'',
math-ph/9903028, DESY, Hamburg, 1999.

\refno\Grigore.
D. R. Grigore,
``On the uniqueness of the non-abelian gauge theories in the
Epstein--Glaser approach to renormalization theory'',
hep-th/9806244, Bucharest, 1998.

\refno\HurthS.
T. Hurth and K. Skenderis,
``The quantum Noether condition in terms of interacting fields'',
hep-th/9811231, 1998.

\refno\Pinter.
G. Pinter,
``The action principle in Epstein--Glaser renormalization and
renormalization of the $\Sb$-matrix of $\Phi^4$-theory'',
hep-th/9911063, DESY, Hamburg, 1999.

\refno\EpsteinG.
H. Epstein and V. Glaser,
``The role of locality in perturbation theory'',
Ann. Inst. Henri Poincar\'e {\bf A XIX} (1973), 211--295.

\refno\Philippe.
P. Blanchard and R. S\'en\'eor,
``Green's functions for theories with massless particles (in
perturbation theory)'',
Ann. Inst. Henri Poincar\'e {\bf A XXIII} (1975), 147--209.

\refno\Scharf.
G. Scharf,
\textit{Finite Quantum Electrodynamics: The Causal Approach},
2nd edition,
Sprin\-ger, Berlin, 1995.

\refno\Polaris.
J. M. Gracia-Bond\'{\i}a, J. C. V\'arilly and H. Figueroa,
\textit{Elements of Noncommutative Geometry},
Birkh\"auser, Boston, 2000.

\refno\Rhea.
J. M. Gracia-Bond\'{\i}a and J. C. V\'arilly,
``QED in external fields from the spin representation'',
J. Math. Phys. {\bf 35} (1994), 3340--3367.

\refno\ShaleS.
D. Shale and W. F. Stinespring,
``Spinor representations of infinite orthogonal groups'',
J. Math. Mech. {\bf 14} (1965), 315--322.

\refno\BogoliubovS.
N. N. Bogoliubov and D. V. Shirkov,
\textit{Introduction to the Theory of Quantized Fields},
Wiley, New York, 1959.

\refno\GrosseL.
H. Grosse and E. Langmann,
``A super-version of quasi-free second quantization: I. Charged
particles'',
J. Math. Phys. {\bf 33} (1992), 1032--1046.

\refno\Ottesen.
J. T. Ottesen,
\textit{Infinite Dimensional Groups and Algebras in Quantum Physics},
Lecture Notes in Physics: Monographs {\bf m27},
Springer, Berlin, 1995.

\refno\LangmannCocycle.
E. Langmann,
``Cocycles for boson and fermion Bogoliubov transformations'',
J. Math. Phys. {\bf 35} (1994), 96--112.

\refno\Aldebaran.
J. C. V\'arilly and J. M. Gracia--Bond\'{\i}a,
``$S$-matrix from the metaplectic representation'',
Mod. Phys. Lett. A {\bf 7} (1992), 659--667.

\refno\Scheck.
F. Scheck,
\textit{Electroweak and Strong Interactions},
Springer, Heidelberg, 1996.

\refno\Estrada.
R. Estrada,
``Regularization of distributions'',
Int. J. Math. \& Math. Sci. {\bf 21} (1998), 625--636.

\refno\BerestetskiiLP.
V. B. Berestetskii, E. M. Lifshitz and I. P. Pitaevskii,
\textit{Quantum Electrodynamics}
(volume~4 of the Landau and Lifshitz Course of Theoretical Physics),
Pergamon Press, Oxford, 1980.

\refno\ItzyksonZ.
C. Itzykson and J.-B. Zuber,
\textit{Quantum Field Theory},
McGraw-Hill, New York, 1980.

\refno\GreinerQED.
W. Greiner and J. Reinhardt,
\textit{Quantum Electrodynamics},
Springer, Berlin, 1992.

\refno\LangmannM.
E. Langmann and J. Mickelsson,
``Scattering matrix in external field problems'',
J. Math. Phys. {\bf 37} (1996), 3933--3953.

\refno\JauchR.
J. M. Jauch and F. Rohrlich,
\textit{The Theory of Photons and Electrons},
Springer, Berlin, 1976.

\refno\Ruijsenaars.
S. N. M. Ruijsenaars,
``Charged particles in external fields. II. The quantized Dirac and
Klein--Gordon theories'',
Commun. Math. Phys. {\bf 52} (1977), 267--294.

\refno\ScharfW.
G. Scharf and W. F. Wreszinski,
``The causal phase in Quantum Electrodynamics'',
Nuovo Cim. A {\bf 93} (1986), 1--27.

\refno\Bellissard.
J. Bellissard,
``Quantized fields in interaction with external fields
I. Exact solutions and perturbative expansions'',
Commun. Math. Phys. {\bf 41} (1975), 235--266.

\bye